\documentclass[manuscript]{aastex6}
\usepackage{epstopdf}

\newcommand{\polyorder}{5th}
\newcommand{\iof}{\frac{\rm I}{\rm F}}

\def\equationautorefname~#1\null{%
  equation~(#1)\null
}

\shortauthors{Mayorga, L. C. et al.}
\shorttitle{Jupiter's \replaced{Phase Variations}{Empirical Phase Curve}}
\begin{document}

\title{Jupiter's Phase Variations from Cassini: a testbed for future direct-imaging missions}
\author{L.~C. Mayorga\altaffilmark{1}, J. Jackiewicz}
\affil{Department of Astronomy, New Mexico State University, Las Cruces, NM 88003-8001, USA}
\author{K. Rages}
\affil{SETI Institute, Mountain View, CA 94043, USA}
\author{R.~A. West}
\affil{Jet Propulsion Laboratory, California Institute of Technology, Pasadena, CA 91109, USA}
\author{B. Knowles}
\affil{CICLOPS/Space Science Institute, Boulder, CO 80301, USA}
\author{N. Lewis}
\affil{Space Telescope Science Institute, Baltimore, MD, 21218, USA}
\affil{Department of Earth and Planetary Sciences, Johns Hopkins University, Baltimore, MD, 21218, USA}
\and
\author{M.~S. Marley}
\affil{NASA Ames Research Center, Moffett Field, CA 94035, USA}
\altaffiltext{1}{NSF GRFP Fellow}

\begin{abstract}
We present \added{empirical} phase curves of Jupiter from \replaced{0--150}{$\sim0-140$}~degrees as measured in multiple optical bandpasses by Cassini/ISS during the Millennium flyby of Jupiter in late 2000 to early 2001. Phase curves are of interest for studying the energy balance of Jupiter and understanding the scattering behavior of Jupiter as an exoplanet analog. We find that Jupiter is significantly darker at partial phases than an idealized Lambertian planet by roughly 25\% and is not well fit by Jupiter-like exoplanet atmospheric models across all wavelengths. \added{We provide analytic fits to Jupiter's phase function in several Cassini/ISS imaging filter bandpasses}. In addition, these observations show that Jupiter's color is more variable with phase angle than predicted by models. Therefore, the color of even a near Jupiter-twin planet \added{observed at a partial phase} cannot be assumed to be comparable to that of Jupiter at full phase. We discuss how WFIRST and other future direct-imaging missions can enhance the study of cool giants.
\end{abstract}

\section{Introduction}
\replaced{The reflectivity of a planet, the ratio of the energy incident on a planet and the energy emitted by a planet, defines its energy budget, barring internal energy sources. The energy budget influences}{The ratio of the total energy scattered by a planet to that incident upon the planet, or the Bond albedo, defines its external energy budget. Along with internal heat flow, this influences.} a planet's atmospheric temperature, controls its evolution through time, and defines its detectability as a function of wavelength. A more reflective atmosphere allows a planet to cool more quickly than a darker, more absorptive atmosphere. Complete measurements of the reflectivity of the planet require observations of the light from the planet relative to the light from the star at all wavelengths at all phase angles, $\alpha$, the angle formed between the star, planet, and Earth. \replaced{These observations are called phase curves and are}{The variation of reflectivity at a given wavelength as a function of phase angle is known as the phase curve and is} defined as follows:
\begin{equation}
\frac{F_p(\lambda, \alpha)}{F_\odot(\lambda)} = A_g(\lambda)\left(\frac{R_p}{d}\right)^2 \Phi(\lambda, \alpha),
\end{equation}
where $A_g(\lambda)$ is the monochromatic geometric albedo defined at $\alpha = 0^\circ$, $F_p(\lambda, \alpha)$ is the monochromatic planet flux \added{observed at phase angle, $\alpha$}, $F_\odot(\lambda)$ is the monochromatic incident solar flux, \deleted{or stellar flux for an exoplanet} $R_p$ is the planet's radius, d is the planet-star separation, and $\Phi(\lambda, \alpha)$ is the planet's phase function at angle $\alpha$, not to be confused with particle scattering phase functions.

The simplest model planetary phase function treats planets as Lambertian reflectors where the reflected intensity is described by
\begin{equation}
\Phi(\alpha) = \frac{\sin(\alpha)+(\pi-\alpha)\cos(\alpha)}{\pi}.
\end{equation}
\replaced{More recent works have simulated phase functions beyond Lambert scattering, considering more}{Other phase functions are possible, including those which arise from} isotropic or Rayleigh scattering \added{(e.g. \citet{Dlugach1974, Greco2015})} or a combination of these and non-isotropic Mie scatterers \citep{Cahoy2010a}. From the phase function, we can calculate the phase integral, 
\begin{equation}
q(\lambda) = 2 \int_0^\pi\Phi(\lambda, \alpha)\sin(\alpha)d\alpha,
\end{equation}
which \replaced{describes is}{leads to} the fraction of light reflected at all angles known as the spherical albedo, $A_s(\lambda) = q(\lambda)A_g(\lambda)$. From this the full Bond albedo, a measure of the amount of light reflected at all angles and all wavelengths, can be computed as,
\begin{equation}
A_B = \frac{\int_0^\infty A_s(\lambda)F_\odot(\lambda)d\lambda}{\int_0^\infty F_\odot(\lambda)d\lambda}.
\end{equation}
\added{For a given internal heat flow and} under the assumption of thermal equilibrium, the effective temperature of a planet can be estimated, allowing major cloud constituents and atmospheric structure to be modeled. \replaced{Formation}{Evolution} models can then constrain the cooling history.\deleted{to approximate an age}

\added{While these concepts are familiar in planetary sciences, they have been fruitfully applied to exoplanet science as well. As reviewed by \citet{Parmentier2016}, measurements of the phase curves of transiting planets have been used to create atmospheric brightness maps (e.g. \citet{Cowan2008} and \citet{Demory2013}).} \deleted{Non-spectroscopic studies of unresolved planets have demonstrated that even white-light observations can reveal information about the atmospheres of giant planets. It has been postulated that phase functions can be used to create atmospheric brightness maps.} Recent phase curve observations have indicated that planets have variations in cloud coverage \citep{Webber2015, Majeau2012a, Demory2013, Parmentier2016, Sing2016} and deviate from the \added{simple} Lambertian assumption. \deleted{replaced{This assumption may overestimate the reflectivity of planets by a factor of $\sim4$ compared to more physically motivated representations like those of}{The reflectivity of this sample of exoplanets has been observed to deviate substantially from the simplistic Lambertian behavior} Greco2015, Cahoy2010a} \added{In the future, planned space telescopes, such as WFIRST, aim to directly image extrasolar planets in reflected light. The interpretation of such observations will require an understanding of their phase curves \replaced{e.g Nayak et al. submitted}{\citep[e.g.][]{Cahoy2010a, Greco2015}} in order to interpret cloud properties and the abundance of various absorbers \added{(e.g., Nayak et al., submitted).}}

\replaced{Unlike exoplanets, the Solar System planets are resolved and we can directly measure atmospheric properties.}{Fortunately, Solar System giant planets are generally well characterized and offer a point of comparison to future exoplanet observations.} However, studies of the outer Solar System planets are complicated by observational limitations. Since the Earth is interior to the giant planets in the Solar System, only phase angles $\lesssim12^\circ$ are visible to us from the ground for Jupiter \added{\citep{Gussow1929, Guthnick1920a, Guthnick1918, Muller1893, Irvine1968}}, and even more restricted phase angles \added{are available} for the more distant ice giants. Observations of full phase curves for the outer planets requires space-craft observations, e.g., \cite{Pollack1986}. The most notable ground-based work \citep{Karkoschka1994, Karkoschka1998} was able to obtain albedo spectra for all the giant planets at low phase angles. These studies have become the basis for many \replaced{space-based works}{exoplanet analyses}. In particular, studies of Jupiter are valuable as both a planet in our Solar System and an exoplanet analog.

\added{In addition to their role in the interpretation of future observations, phase curves also play a role in mission modeling and observation planning. The brightness of planets as a function of phase is an important input parameter to simulations of exoplanet searches by direct imaging and observation planning for known planets detected by other means. While it has been known since the early 20th century that Jupiter is not Lambertian, only recently have non-Lambertian phase curves been considered in the exoplanet arena. To aid in such efforts here we seek to provide a testbed specifically for giant-planet characterization and searches with direct-imaging techniques. Beyond these empirical phase curves, the literature of the characterization of giant planets, including such works as \citet{Tomasko1978a} and \citet{Smith1984a}, will undoubtedly be helpful for exoplanet direct imaging science.}

The need to classify the type of object being observed has led astronomers to often turn to color. Color-color diagrams have been discussed as a method of determining surface properties of terrestrial-type planets \citep{Hegde2013}. Also, the color differences between objects have opened a possibility for detecting exomoons through spectroastrometric observations \citep{Agol2015} and can be used to mitigate the contamination from background objects \citetext{Exo-S Final Report}. In particular, \citet{Cahoy2010a} discuss the use of color-color diagrams to help distinguish between Neptune-like and Jupiter-like exoplanets. In stellar astronomy, the $(U - B)$, $(B - V)$ colors have a long history. The $(U - B)$ index roughly measures the height of the Balmer jump and has been shown to be sensitive to both surface gravity and effective temperature, with further complications in interpretation arising from a sensitivity to metallicity. The $(B - V)$ index is sensitive to the slope of the Paschen continuum and is correlated with the effective temperature of the star. The choice of photometric system has since somewhat evolved to more precisely measure stellar features. 

From the combination of spectra compiled by \citet{Karkoschka1994, Karkoschka1998}, \citet{Sing2016}, and \replaced{the modelling work of Cahoy2010}{modelling \citep[e.g.,][]{Marley1999,Sudarsky2000, Cahoy2010a}}, we can see that there are several notable features that would help differentiate between \added{various classes of gas giant} exoplanets. \replaced{The blue end slope is a product of the optical scattering of hazes. The features deep into the red and near-infrared are mainly caused by methane with variations between Jupiter and Uranus resulting from differences in ammonia, water, and alkali metal abundances.}{In general the blue end slope is a product of Rayleigh scattering by gas and hazes. Into the red and near-infrared are spectral features of methane, ammonia, water, and alkali metals. Differences among solar system giants are primarily attributable to variations in gravity, cloud properties, and methane abundance.}

The appropriate selection of color indices for best identifying giant planets is \replaced{under much discussion}{relevant to instrument filter selection and mission planning for direct imaging campaigns}. \added{For example} among all the combinations of Johnson-Morgan/Cousins (JC) filters, \citet{Cahoy2010a} found that the $(B - V)$, $(R - I)$ and $(B - V)$, $(V - I)$ filter pairs showed the greatest spread between planet types when considering how color evolves with phase angle. To test \replaced{these Jupiter-like planet predictions}{the veracity of these exoplanet models} \added{for the Jupiter-like planet case} we require space-based observations across many phase angles and multiple bandpasses.

\replaced{The latest space-based work resulting in}{Prior to the recent arrival of the Juno Spacecraft to Jupiter, the latest} full-disk imaging of Jupiter came in 2000, in a flyby of the planet by Cassini dubbed the ``Millennium'' flyby. Cassini was \replaced{intended}{en route} to orbit around Saturn and carried a suite of instruments designed to study the atmospheres of Saturn and Titan, the particles in the rings, and more \citep{Porco2003}. \deleted{En route to Saturn,}The flyby provided thousands of full-disk images of Jupiter in a variety of filters. An in depth explanation of the cameras and filters on the \replaced{ISS}{Imaging Science Subsystem (ISS)} can be found in \cite{Porco2005} and is briefly discussed in \autoref{sec:data}.

Previous attempts to measure Jupiter's phase curves used data from Voyager \citep{Hanel1981} and Pioneer 10 and 11 \citep{Tomasko1978a, Smith1984a}, but they had \added{either} limited phase coverage, wavelength coverage, or were lacking in full-disk measurements at all phases\added{, which are most relevant to direct-imaging studies of exoplanets}. Cassini/ISS data has already provided information about the atmospheric properties of Jupiter \citep{Zhang2013, Sato2013}, but previous studies have been limited to the Narrow Angle Camera (NAC), one of the two components of the ISS, in concert with older space-based or ground-based observations \citep{Dyudina2005, Dyudina2016}. 

Future exoplanet missions have prioritized atmospheric characterization via direct-imaging. The Wide-Field InfraRed Survey Telescope (WFIRST) was identified as the top priority space mission in the 2010 ``New Worlds, New Horizons'' decadal survey. Its current design \citep{WFIRST} features a coronagraph instrument (CGI) with an inner working angle of $\sim0.1$~arcseconds. The CGI will enable WFIRST to directly image the light from exoplanets at moderate phase angles near quadrature. The design features imaging capabilities in visible wavelengths (430--970~nm) with detection limitations reaching unprecedented planet/star contrast ratios \citep[see Figure 2-44 in][]{WFIRST}.

Preliminary target lists, sourced from planets detected through the radial-velocity measurements, demonstrate that WFIRST's most promising targets will be Jupiter-mass planets with semi-major axes of only a couple AU \citep[Table 2-7 in][]{WFIRST}. These planets maintain high contrast ratios due to their size and proximity to their host stars without being within the inner working angle. In preparation for the analysis of these \deleted{unresolved} systems, it is critical that we gain an understanding of how Jupiter-like planets scatter light as function of phase angle because they cannot be directly imaged at $0^\circ$ phase. The Cassini/ISS instrument shares the same wavelength regime and has provided nearly full phase angle coverage of Jupiter and is therefore perfectly poised to provide observations of Jupiter of the type WFIRST will conduct on giant exoplanets.

We present here an analysis of the images from both ISS cameras, the Narrow Angle and Wide Angle Camera (NAC and WAC), in six filters ranging from roughly 400--1000~nm and spanning phase angles from about \replaced{0--150}{0--140} degrees. In \autoref{sec:data} we present our data, the selection process (\autoref{subsec:sel}), the reduction software (\autoref{subsec:reduc}), and photometric analysis process (\autoref{subsec:proc}, \autoref{subsec:phot}). In \autoref{sec:results} we present the phase functions of Jupiter. We discuss our results and implications for observations and model performance in \autoref{sec:disc} while making predictions for WFIRST. We close in \autoref{sec:conc} with a summary of our results and preparations for future missions.

\section{Data and Reduction}
\label{sec:data}
\begin{figure}
	\centering
	\includegraphics[width=\columnwidth]{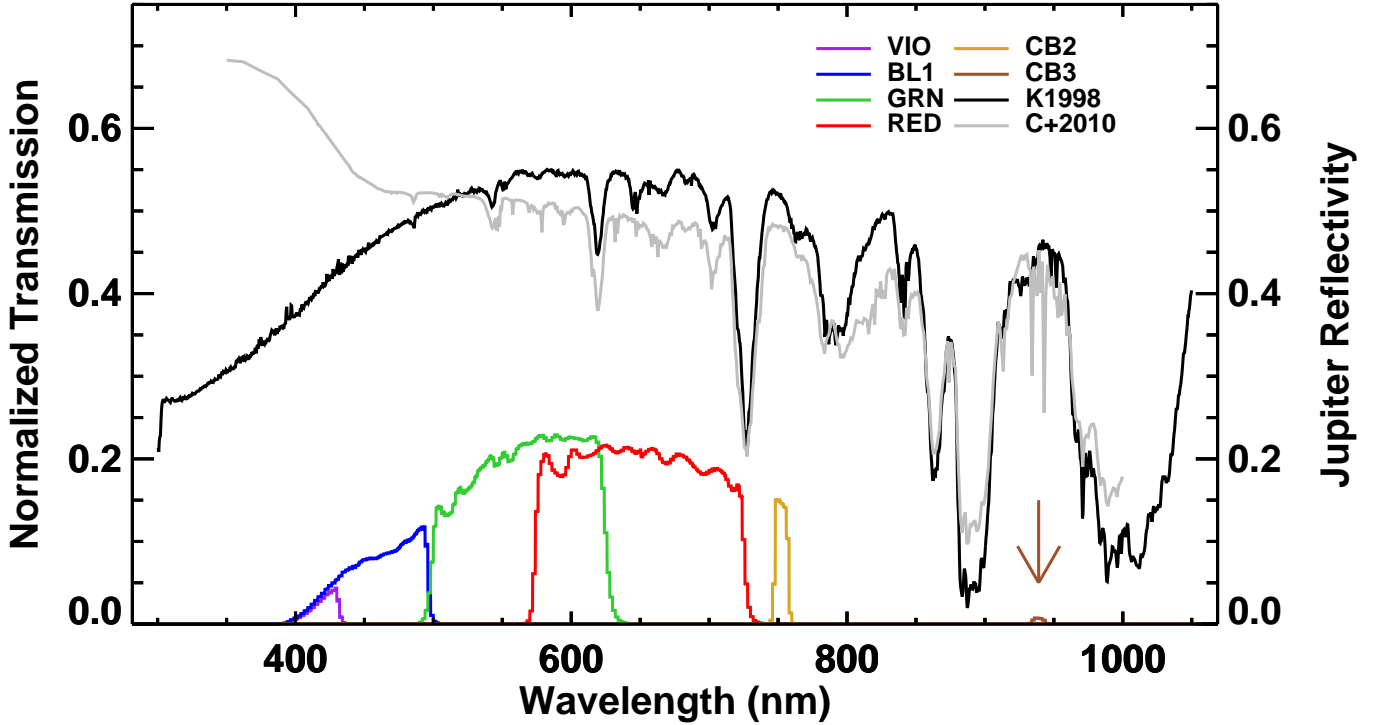}
	\caption{The filter transmission curves for the WAC filters used in this study. Overplotted are the data from \citet{Karkoschka1998} taken in 1995 at a phase angle of $6.8^\circ$ and the \citet{Cahoy2010a} model of a Jupiter-like planet at a phase angle of $0^\circ$ degrees with no photochemical hazes. \label{fig:trans}}
\end{figure}

Cassini/ISS took almost 23,000 images of Jupiter during the Millennium flyby spanning from October, 2000 to March, 2001. Cassini imaged Jupiter with both the WAC and the NAC, with fields of view of $3.5^\circ$ and $0.35^\circ$, respectively \citep{Porco2005}. The two cameras each have two filter wheels that are designed to move independently. The WAC has 9 filters on each wheel for a total of 18 filters. The NAC has 12 filters on each wheel for a total of 24 filters. Roughly half of the NAC filters are duplicated on the WAC, however the bandpasses are shifted a few nm because of the differing spectral transmissions of the camera optics (the optical train of the WAC is a Voyager flight spare), \citep{Porco2005}. These 15 common filters span from 400--1000~nm in wavelength and six of the WAC filters' transmission curves are shown in \autoref{fig:trans}.

\subsection{Data Selection}
\label{subsec:sel}
\begin{figure}
	\centering
	\includegraphics[width=\columnwidth]{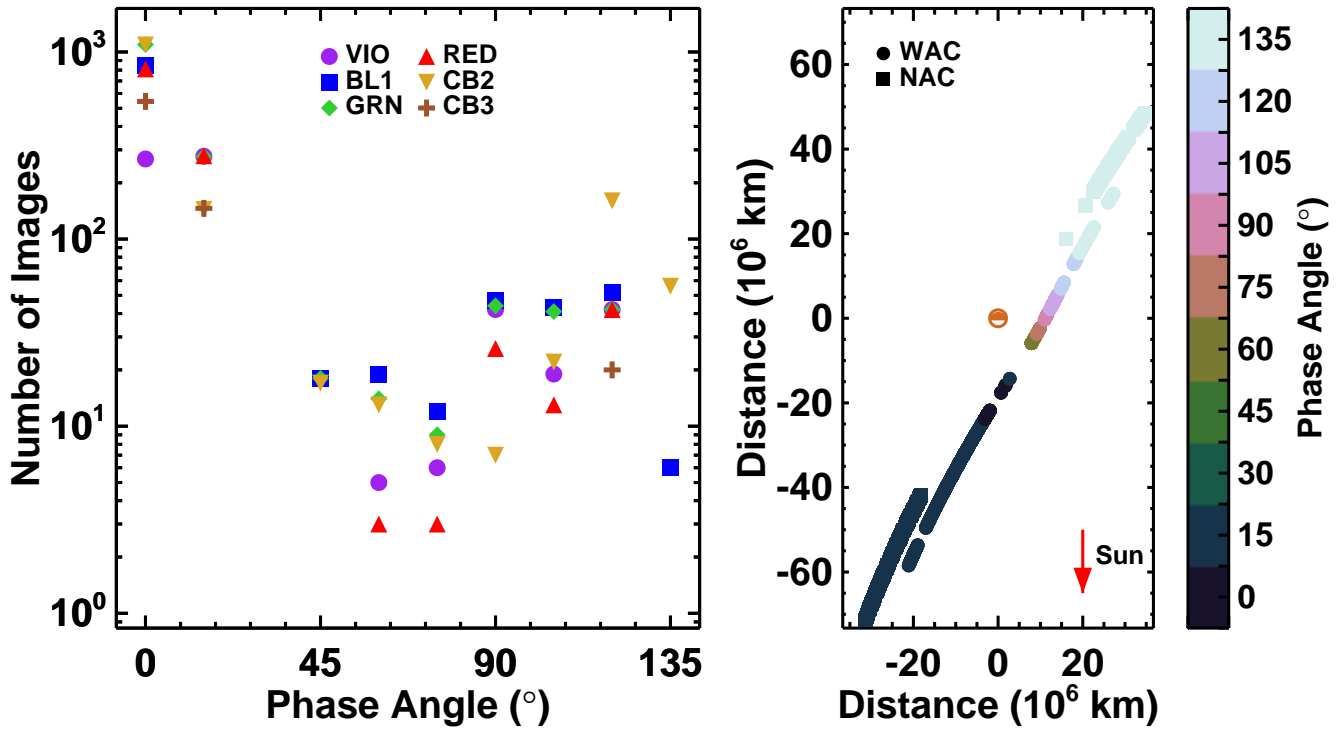}
	\caption{The phase angle coverage available in a selection of filters for both the WAC and NAC cameras. \added{Left:} The number of images that fit our selection criteria is shown for \replaced{5 degree}{$15^\circ$} bins in phase angle. The gap in coverage from 20 to 50 degrees is due to Jupiter encompassing an area too large to be contained in a single image \added{as well as the spacecraft targeting for other observations. Right: The position of Cassini around Jupiter during the flyby in the J2000 $(x,y)$ plane color-coded with phase angle in $15^\circ$ bins. For clarity, the NAC data (squares) are offset from the WAC data (circles) by $-5$ million km in $x$.}.\label{fig:coverage}}
\end{figure}

The images vary in resolution and orientation during the course of the flyby. The requirement for this study is that Jupiter be contained entirely within an image. \deleted{We chose to constrain Jupiter to subtend less than {\onethird} of the field of view in order to capture all of the light spread by the point spread function (PSF) as well as have sufficient background for accurate subtraction. This constrains the pixel scale of any image to be larger than $200\,\rm km\,pixel^{-1}$.} \added{This requirement resulted in the poor sampling of phase angles in the 20--50 degree range (see \autoref{fig:coverage}), where not only was Jupiter too large, but the spacecraft also made other observations. During the photometry process, we ensure that Jupiter's disk does not fall off the edge of the field-of-view due to Jupiter being too close or from specific pointings.}

Given the need for short exposures to prevent smearing during flybys, the most common \replaced{set up}{configuration} is a clear filter in combination with a filter from the other wheel. For the WAC specifically, the CL1 filter was designed to improve focus when combined with the filters of the second wheel. Sharp focus is difficult to achieve on the WAC with any other combination and renders most two-filter combinations useless \citep{Porco2005}. Maximizing the phase coverage of the data results in the bulk of the images being a combination of a clear filter and a common filter that is present on the WAC's second wheel.

Of the 15 similar filters on the WAC and NAC, only data observed with six of them in combination with a clear filter are considered in this study (the six shown in \autoref{fig:trans}). We discarded the two methane-band filters (MT), the four IR filters, an IR polarization filter, and a Hydrogen alpha filter. These 8 filters do not have sufficient phase coverage for our purposes. We opted to include the VIO filter, despite it not being a common filter, because it does have sufficient phase coverage. Due to the methods by which the data were transmitted to Earth, those images that were compressed in a lossy fashion were also discarded as they are not of photometric quality.


This selection process results in nearly 10,000 images being selected and downloaded from the Planetary Data System (PDS) Rings Node in the raw format. Some additional images were removed from our sample due to pixel saturation, poor targeting of Jupiter, etc. The final sample is roughly \replaced{5000}{6600} images. The resulting number of images as a function of phase angle for the six filters is shown in \autoref{fig:coverage}. The total number of images and effective wavelengths of the filters use in this study, computed from using the full system transmission function convolved with a solar spectrum, are shown in \autoref{tbl:counts}.

\floattable
\begin{deluxetable}{lrrrr}
\tablecolumns{5}
\tablewidth{0pt}
\tablecaption{\label{tbl:counts} Number of images in each filter used in this study.}
\tablehead{
\colhead{Filter}\hspace{1em} & \multicolumn{2}{c}{WAC} & \multicolumn{2}{c}{NAC} \\
\colhead{} & \colhead{$\lambda_{\rm eff}$ (nm)} & \colhead{\# of images} & \colhead{$\lambda_{\rm eff}$ (nm)} & \colhead{\# of images}}
\startdata
VIO & 420 & 658 & \nodata & \nodata \\
BL1 & 463 & 201 & 455 & \replaced{426}{841} \\
GRN & 568 & 701 & 569 &  \replaced{429}{839} \\
RED & 647 & 622 & 649 & \replaced{147}{556} \\
CB2 & 752 & 680 & 750 & \replaced{424}{837} \\
CB3 & 939 & 421 & 938 & \replaced{288}{289} \\
\hline
Total & \nodata & 3283 & \nodata & \replaced{1714}{3362}\\
\enddata
\end{deluxetable}

\subsection{Cassini ISS CALibration (CISSCAL) image reduction}
\label{subsec:reduc}
CISSCAL is the Cassini/ISS image reduction pipeline. CISSCALv3.6 was the latest version at the onset of our study and we followed the steps outlined by \citet{West2010} and the ISS Data User Guide\footnote{pds-rings.seti.org/cassini/iss/ISS\_Data\_User\_Guide\_120703.pdf} \added{to fully reduce the data}. To summarize, CISSCAL removes bias, dark current, and a 2~Hz noise source found in all ISS data. It applies flatfielding, nonlinearity, and antiblooming corrections and accounts for exposure times, quantum efficiency, and uneven bit weighting. The images resulting from this process are provided in photon flux units alone or normalized by the solar flux ($\iof$). The latter case will be denoted as the reflectivity. 

\begin{figure}
	\centering
	\includegraphics[width=\columnwidth]{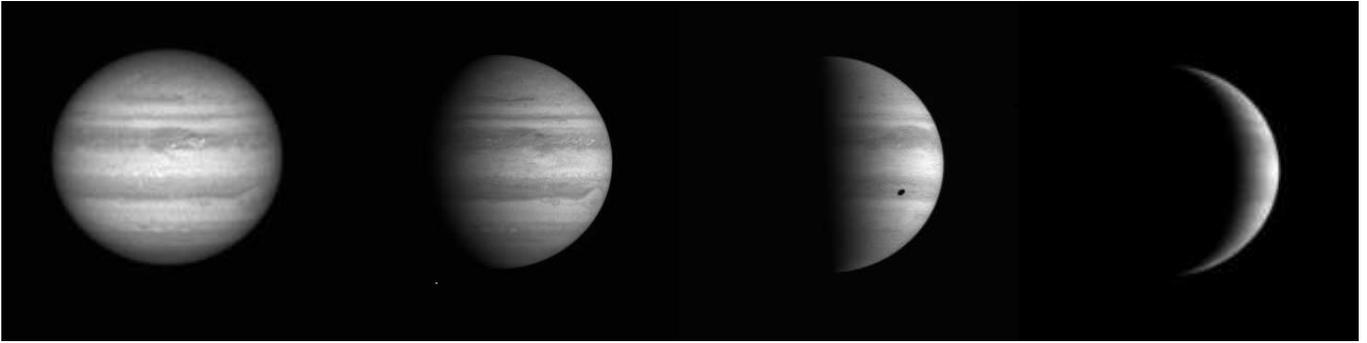}
	\caption{Example images of Jupiter taken in the GRN filter after CISSCAL reduction. The flyby allowed Cassini to image Jupiter at various phases shown here from left to right at approximately 0, 50, 90, and 125 degrees. Note that the 90 degree image contains the shadow of the moon Io. \label{fig:phase}}
\end{figure}

Reduced images still vary in resolution, orientation, and presence or absence of moons and their shadows on Jupiter's disk. A sample of four fully reduced green (GRN) WAC images from crescent to full phase that have been properly derotated, aligned, and cropped is shown in \autoref{fig:phase}.

\subsection{Additional data processing}
\label{subsec:proc}
This study only concerns Jupiter's reflected and scattered light and the fully reduced images still require removal of some wayward light sources. These sources can be cosmic rays, background stars, and moonlight. We discuss \added{lost} light which has not made it to the detector due to PSF broadening. It is also necessary to adjust for the non-equatorial viewing angle \replaced{reducing}{increasing} Jupiter's disk area.

The corrections for lost light and cosmic rays are described by \citet{West2010}. We chose to forgo the lost light correction partially due to analysis in \citet{West2010}, who note it made less than a 1\% difference in the majority of filters. In addition, the PSFs were derived from observations of stars in early flight stages. Due to the method that the PSFs were created, it is suspected that the wings of the extended PSFs are overestimated and require attenuation from their current values. Our analysis of the Jupiter images, as \citet{West2010} show for Enceladus, gives us confidence that indeed the lost light fraction can be ignored at the 1\% level.

Jupiter's apparent size and the position of its moons vary on an image-to-image basis. We required navigation and targeting information for each image. We used SPICE (Spacecraft Planet Instrument Camera matrix Events), a program provided and maintained by JPL and acquired from the PDS Navigation Node. The purpose of SPICE is to track observation geometry and events leading to determining the following: (1) spacecraft location and orientation; (2) target location, shape, and size; (3) events on the spacecraft or the ground that might affect the interpretation of science observations \citetext{SPICE overview tutorial 3}. SPICE can quickly determine the 3-dimensional position of Jupiter and its moons, and compute the location of moon shadows on Jupiter's disk from any perspective.

Visual inspection of the images reveals that we often see Io or its shadow crossing Jupiter's disk. The targeting information shows that one of the Galilean satellites and several smaller moons like Adrastea and Metis are often present (see Io's shadow in \autoref{fig:phase} in the $90^\circ$ phase angle image). SPICE, in combination with the image labels, is able to provide us with the necessary position information to identify moons and their shadows that impact the light the cameras receive from Jupiter. At most their presence could cause a \replaced{$\lesssim4$\%}{$\lesssim0.15$\%} difference in the measured flux\added{, (assuming Jupiter is 100\% reflective and the moons are 0\% reflective)}. In the spirit of what will be done with WFIRST, we have chosen to make analogous observations by not removing or correcting for the moons and their shadows in the image, since direct-imaging missions will not be able to resolve such details.

\subsection{Photometry}
\label{subsec:phot}
We briefly discuss how the brightness of Jupiter (relative to the incoming solar flux) in each filter is obtained from the fully-reduced images. To remove leftover background features, such as stellar and lunar sources outside of Jupiter's disk, we first set pixel values that are below a threshold, that is image and filter dependent, to zero. We then determine the location of Jupiter in the image and remove any sources of light that remain outside of Jupiter's disk. What remains is Jupiter's light, some moonlight from moons in front of Jupiter, and the surrounding glow from scattered light. The images are then summed over and normalized by the apparent area of Jupiter to create a phase curve according to
\begin{equation}
\label{eq:norm}
\frac{F_p(\lambda, \alpha)}{F_\odot(\lambda, \alpha)} = \frac{\sum_i^N \iof_i}{A_J},
\end{equation}
where $\iof$ is the solar-flux normalized brightness of Jupiter in a given bandpass at a given phase angle, $A_J$ is the apparent area of Jupiter in pixels, and the sum runs over all $N$ pixels in an image.

Note that the phase curve formulation already accounts for Cassini's distance from Jupiter due to how $A_J$ is computed. \replaced{It is critical that we}{We chose to} properly compute $A_J$, instead of removing a distance factor from $\iof$, because Jupiter is an oblate spheroid. Its apparent shape and area changes from image to image. SPICE is able to compute the location of the apparent limb of Jupiter given Cassini's and Jupiter's 3-dimensional location in space. From this we determine the pixel location of the limb and compute the apparent area $A_J$ in each of our images. Each image has a known phase angle and the photometry of all images produces a phase curve. Normalizing the phase curve to unity at zero phase produces a \emph{phase function}.

\added{The reflectivity values for the NAC and the WAC are offset from each other in each filter. This is likely due to a contamination event and subsequent de-contamination cycle that occurred between the Jupiter and Saturn encounters that changed the calibration. To derive the normalized phase functions, which make use of both datasets, the NAC reflectivity values are first scaled to the WAC reflectivity values. Then, the combined phase curve is fit by a low order polynomial to approximate a $0^\circ$ phase value and normalized to unity. This is easily done for all but the RED filter where there is no phase coverage overlap between the NAC and the WAC values. We instead found the closest NAC and WAC phase angle pair, those closest to $135^\circ$ phase, and scaled based on those. The closest WAC data point is at a lower phase angle ($\sim120^\circ$) while the closest NAC data point is at a higher phase angle ($\sim137^\circ$). The GRN and CB2 filters show that between $120^\circ$ and $137^\circ$ the reflectivity of Jupiter decreased 63\% and 65\%, respectively. Thus we scaled the NAC RED reflectivity values to match a 65\% falloff.}

\section{Results}
\label{sec:results}
\begin{figure}
\centering
\includegraphics[width=\columnwidth]{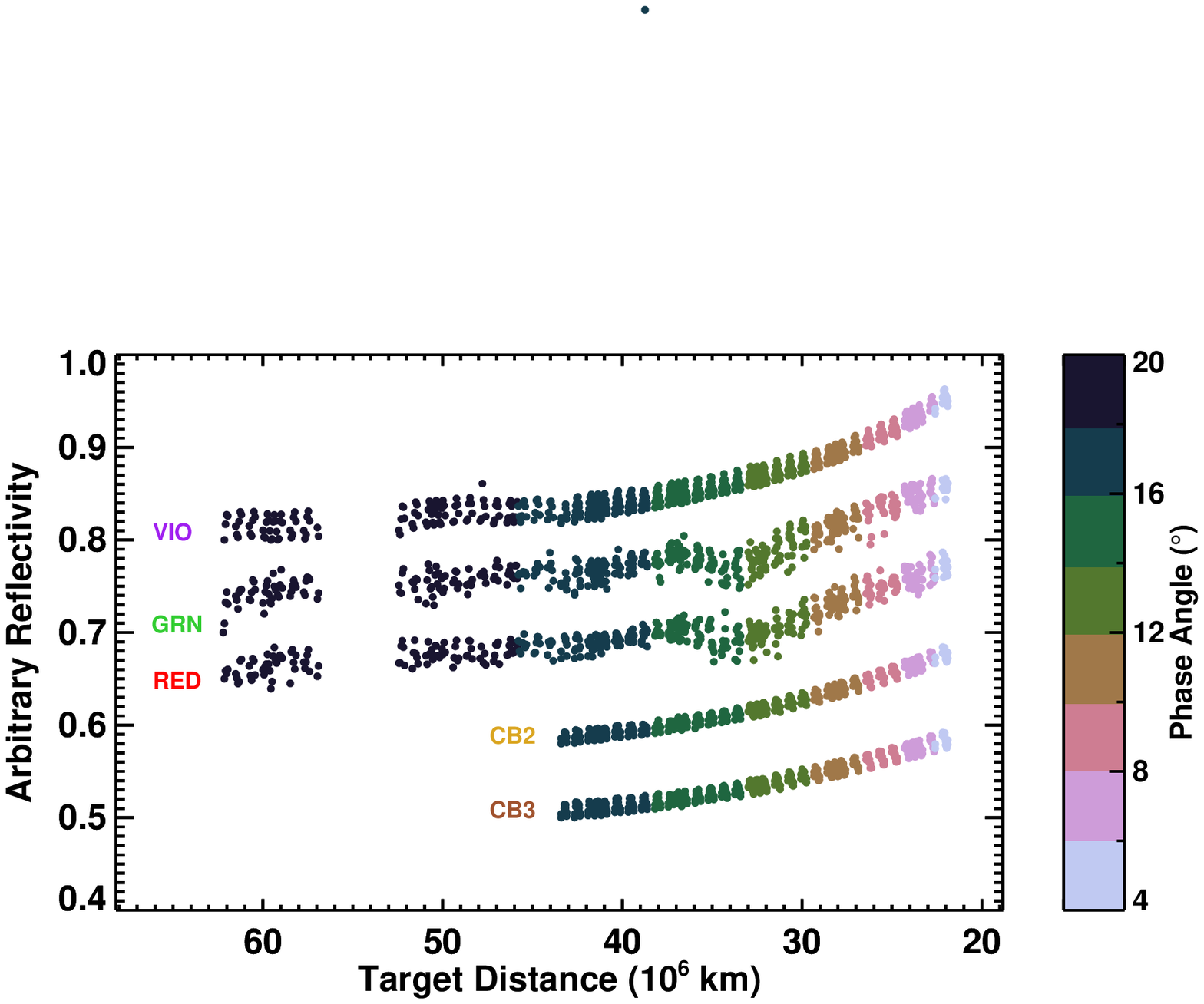}
\caption{\label{fig:stability} The measured reflectivity of Jupiter in 5 of the WAC filters as a function of distance during the period of approach. The deviation seen in the WAC GRN and RED filters is not present in any of the other filters with concurrent imaging. The curves are offset for clarity.}
\end{figure}

During Cassini's approach to Jupiter, the measured flux by the WAC shows peculiar deviations only in certain filters at distances of about 30--35 million~km, corresponding to phases 12--16 degrees. \autoref{fig:stability} shows 5 of the WAC filters during the period of interest. Although there is also higher dispersion in the data at low phase angles, the deviation is markedly present in the GRN and RED filters, but not present concurrently in the VIO, CB2 or CB3 filters. BL1 images were not taken during this time period with sufficient frequency to note such a trend. The NAC does not have images closer than 40 million~km and thus does not cover the period of most deviation. Due to Jupiter's brightness in both the GRN and RED filters, these filters have the shortest exposure duration of 5~ms throughout the entire encounter with Jupiter. Inconsistencies in the shutter speed may be the source of these deviations.

\deleted{The reflectivity values for the NAC and the WAC are offset from each other in each filter. This is likely due to a contamination event and subsequent de-contamination cycle that occurred between the Jupiter and Saturn encounters that changed the calibration. To derive the normalized phase functions, which make use of both datasets, the NAC reflectivity values are first scaled to the WAC reflectivity values. Then, the combined phase curve is fit by a low order polynomial to approximate a $0^\circ$ phase value and normalized to unity. This is easily done for all but the RED filter where there is no phase coverage overlap between the NAC and the WAC values. We instead found the closest NAC and WAC phase angle pair, those closest to $135^\circ$ phase, and scaled based on those. The closest WAC data point is at a lower phase angle ($\sim120^\circ$) while the closest NAC data point is at a higher phase angle ($\sim137^\circ$) and thus this leads to inflated phase function values at high phase angles in the RED. We further scale down the NAC RED reflectivity values by 65\% to match the falloff shown by the GRN and CB2 filters between 120 and 137 degrees (63\% and 65\% respectively).}

\begin{figure*}
\centering
\includegraphics[width=\linewidth]{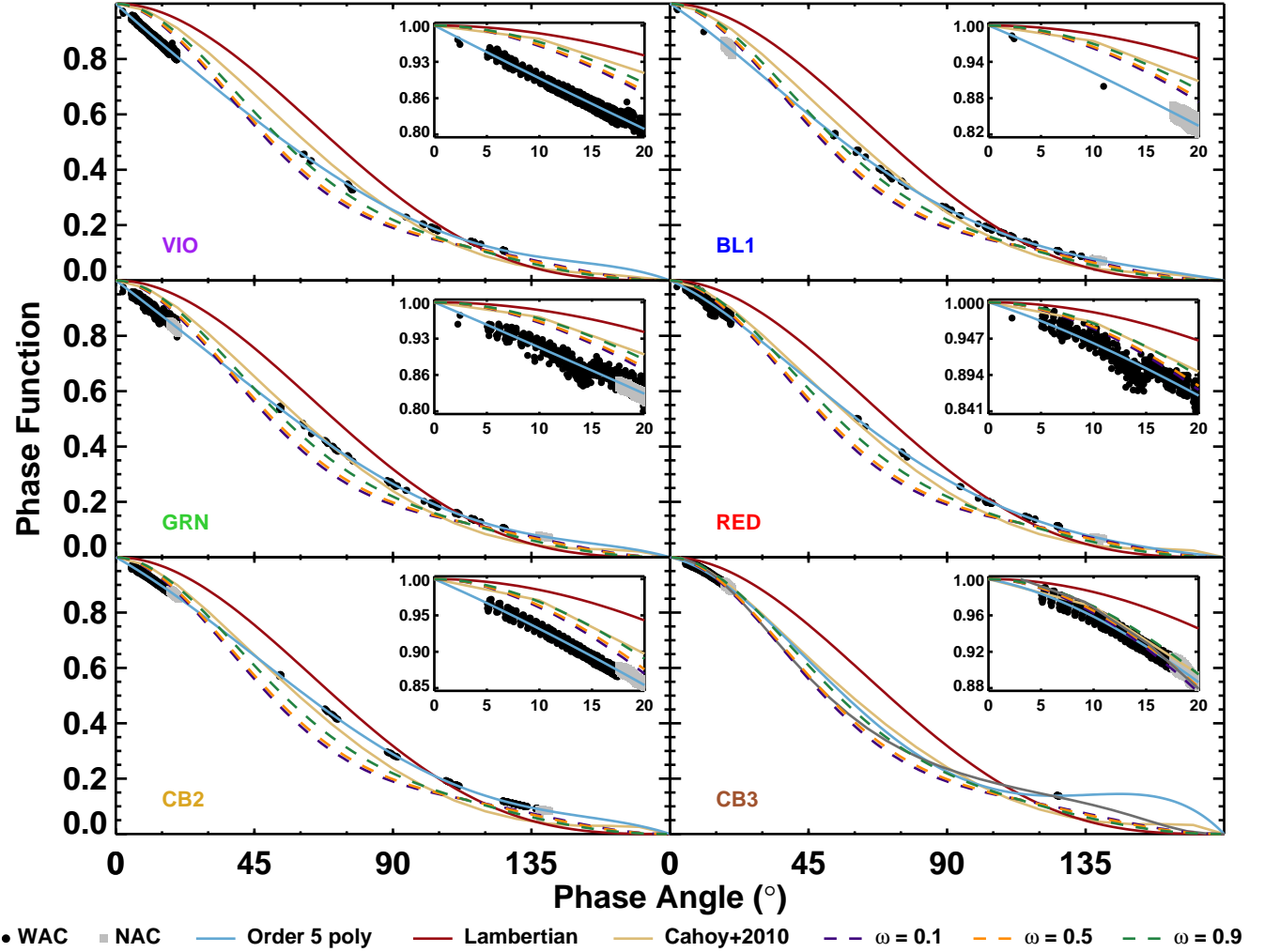}
\caption{\label{fig:models} The phase functions for all of the filters in our study. From the top left to the bottom right we show: VIO, BL1, GRN, RED, CB2, and CB3. The black circles are from the WAC and the gray squares are from the NAC. We overplot a {\polyorder} order polynomial fit for each filter in blue, a Lambertian phase function in red, the phase function derived from filter integrating \citet{Cahoy2010a} in tan, and several \citet{Madhusudhan2012} vector Rayleigh phase functions for a variety of scattering albedos, $\omega$. The \citet{Dyudina2016} phase function for CB3 is overplotted in that panel in dark gray for comparison. The inset plots highlight the region near full phase and demonstrate the steep slope we refer to as the cusp effect.}
\end{figure*}

To compare with the uniformly cloudy Jupiter-like exoplanet from \citet{Cahoy2010a}, we first must generate the phase curves. The \citet{Cahoy2010a} models generated albedo spectra at a $10^\circ$ phase increments. We filter-integrated each albedo spectrum and then normalized to unity at zero phase. The resulting phase functions from the Cassini/ISS and \citet{Cahoy2010a} are shown in \autoref{fig:models}. Also shown are the best fit polynomials to the Cassini/ISS data and a sample of phase functions for a cloud-free atmosphere representing a range of Rayleigh single scattering albedos. 

The phase coverage of NAC images is significantly less than that covered by WAC images, but it does extend coverage into high phase angles. The gap in data in the 20--45 degree phase range, when Cassini was quite close to Jupiter, is caused by the limitations we imposed on Jupiter's apparent size.

The GRN and RED images from the WAC are quite noisy due to the probable shutter instabilities. The other WAC filters and all the NAC filters do not exhibit the same behavior and as a result are much cleaner. However, the VIO images from the WAC show a small bump around $18^\circ$ phase, evident in \autoref{fig:models}, which is not corroborated by any other filters. CB2 images appear to be the cleanest and have the best phase coverage.

We fit the phase functions with a {\polyorder}-order polynomial letting all coefficients be free parameters and restricting the function to reach 0 at $180^\circ$ phase. We also tried a fit by setting the coefficient of the linear term to zero, which fixes the first-derivative to zero at $0^\circ$ phase. The former function yields a better overall fit as the data fall off steeply at low phase angles. Given the phase coverage of the data, the fits break down and are untrustworthy beyond $\sim130^\circ$ due to the lack of data to constrain that region. The table of coefficients of the best fit {\polyorder} order polynomial for each filter is shown in \autoref{tbl:polycoeffs}, where \replaced{$\alpha$ is the phase angle in degrees}{$x=\alpha/180$}. \added{We show the coefficients to the third decimal place to ensure that in the range of 0--135 degrees ($x=$0--0.75), where the fits are trustworthy, the computed curve is within 1\% of the machine precision result.}

\floattable
\begin{deluxetable*}{lrrrrrr}
\tablecolumns{7}
\tablewidth{0pt}
\tablecaption{\label{tbl:polycoeffs} Coefficients of the polynomial fits.}
\tablehead{
\colhead{Filters} & \colhead{[ ]} &\colhead{+\hspace{2em}[ ]\,$x$} & \colhead{+\hspace{2em}[ ]\,$x^2$} & \colhead{+\hspace{2em}[ ]\,$x^3$} & \colhead{+\hspace{2em}[ ]\,$x^4$} & \colhead{+\hspace{2em}[ ]\,$x^5$}}
\startdata
VIO & 1.000 & -1.815 & 0.940 & -2.399 & 4.990 & -2.715 \\
BL1 & 1.000 & -1.311 & -2.382 & 5.893 & -4.046 & 0.846 \\
GRN & 1.000 & -1.507 & -0.363 & -0.062 & 2.809 & -1.876 \\
RED & 1.000 & -0.882 & -3.923 & 8.142 & -5.776 & 1.439 \\
CB2 & 1.000 & -1.121 & -1.720 & 1.776 & 1.757 & -1.691 \\
CB3 & 1.000 & -0.413 & -6.932 & 11.388 & -3.261 & -1.783 \\
\enddata
\end{deluxetable*}

\added{We opted to use a {\polyorder} order polynomial because it minimized the standard deviation of the residuals in each filter (the average standard deviation is 0.003). With increasing order, the average standard deviation is reduced, but individual standard deviations begin to fluctuate. In particular, the lack of constraints in the data for the CB3 filter lead to poorly behaved fits for polynomials greater than 5th order.} \added{Limitations in the available dataset translate into some poorer quality polynomial fits. The VIO filter has no data past $120^\circ$ leading to a poor fit in that region. In addition, the CB3 filter has insufficient mid and high phase angle coverage leading to poor constraints on the fit.}

\added{Furthermore, work by \citet{Karalidi2015} with the Hubble Space Telescope has shown that Jupiter varies by a couple percent in brightness over the course of a rotation and the amplitude of this variation fluctuates with time and the evolution of cloud features. We have presented here only a snapshot of Jupiter in time and fully expect that Jupiter will likewise vary when observed at other phase angles. In addition, observations from higher inclinations will detect smaller phase curve amplitudes and may miss any backscattering anomalies.}

\replaced{The data, as exhibited by the polynomial fits, are clearly inconsistent with a purely Lambertian phase function. We will refer to the uncharacteristic steep slope in the low phase angles as the \emph{cusp effect}. The cusp effect has been readily apparent since the first ground-based observations were conducted at the turn of the 20th Century and most recently was discussed in Dyudina2016 as being caused by large cloud particles sharpening the backscattering peak.}{As seen in Figure 5, the observed phase curves depart from a purely Lambertian phase function and qualitatively more closely resemble the combined Rayleigh and particle scattering phase functions predicted by models \citep{Cahoy2010a, Greco2015}. In general the planet is darker at all wavelengths at phase angles below about 90 degrees and brighter than Lambertian at higher phase angles.} At phase angles above 45 degrees, the Cassini data generally have a shallower slope than a Lambertian phase function. \replaced{The VIO filter has no data past $120^\circ$ leading to a poor fit in that region. In addition, the CB3 filter has insufficient mid and high phase angle coverage leading to poor constraints on the polynomial fit. A more physical representation was discussed by Dyudina2016 through modeling of the Henyey-Greenstein functions and modeling the corresponding light curve. The phase function resulting from this procedure is shown in the CB3 panel of fig:models in dark gray.}{In addition at very low phase angles there is an uncharacteristic steep slope to the phase curve, known as the cusp effect, that has been readily apparent since the first ground-based observations were conducted at the turn of the 20th Century \added{\citep{Gussow1929, Guthnick1920a, Guthnick1918, Muller1893, Irvine1968}} and most recently was discussed in \citet{Dyudina2016} as being caused by large cloud particles sharpening the backscattering peak.  \citet{Dyudina2016} employ models incorporating Henyey-Greenstein functions to study this effect in detail.  The phase function resulting from their modeling is shown in the CB3 panel of \autoref{fig:models} in dark gray.}

We also compared our phase functions with model representations. The first comparison is with the phase function of the 3x solar metallicity Jupiter-like planet at a separation of 5~AU from \citet{Cahoy2010a}. This model\added{\footnote{https://wfirst.ipac.caltech.edu/sims/Exoplanet\_Albedos.html}} is computed to provide an approximate ``Jupiter-like'' albedo spectrum although it is missing the atmospheric photochemistry that would more accurately reproduce the hazes and chromophores that would lower the measured albedo in the blue and UV to match that of Jupiter (see again \autoref{fig:trans}). 

The second comparison is with several phase functions produced through a vector Rayleigh-scattering treatment by \citet{Madhusudhan2012}. Rayleigh scattering is the dominant contribution to scattered light in a cloud-free atmosphere \citep{Madhusudhan2012}. The models have a range of scattering albedos, $\omega$, given by
\begin{equation}
	\omega = \frac{\sigma_{\rm scat}}{\sigma_{\rm scat} + \sigma_{\rm abs}},
\end{equation}
where $\sigma_{\rm scat}$ and $\sigma_{\rm abs}$ denotes the single-scattering and absorption cross sections, respectively. For a perfectly Rayleigh-scattering atmosphere, $\omega = 1$. A more absorptive atmosphere can be modeled with smaller values of $\omega$. The scattering albedo is highly dependent on wavelength and atmospheric conditions such as temperature, pressure, and composition. Thus we have selected a sub-sample of the full range of potential $\omega$ values for comparison.

\section{Discussion}
\label{sec:disc}
WFIRST will be observing planets which are well separated from their host stars. For edge-on systems, the planet will be observed at phase angles somewhere between 40 and 130 degrees depending on the orientation of the orbit and the limitations of the coronographs inner working angle. For systems closer to face-on, this range shrinks down towards $\sim90^\circ$. \replaced{It is critical to understand what is being observed in these near-quadrature regimes}{The need to understand scattering from the planetary atmosphere in these near-quadrature regimes is clear.}

\begin{figure}
\centering
\includegraphics[width=\columnwidth]{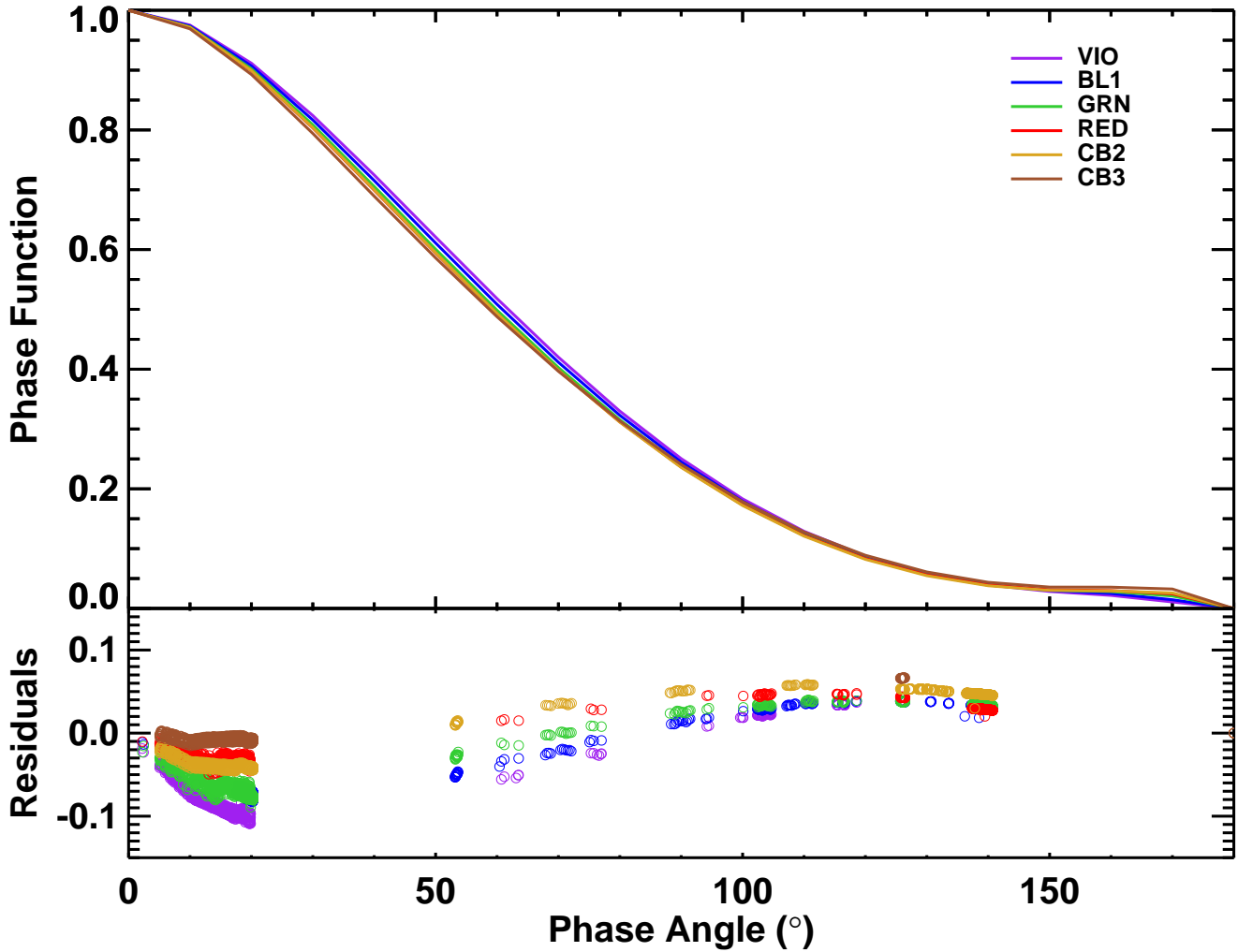}
\caption{\label{fig:cahoy} \added{Upper:} The \citet{Cahoy2010a} phase functions of a 3x solar metallicity Jupiter with a separation of 5~AU from a solar-like star as seen through the Cassini filter set. \added{Lower: The residuals between the data and each model.}}
\end{figure}

The Cassini data indicate that Jupiter's \added{true} phase function \added{is not perfectly reproduced by any current model prediction. Indeed it} differs from the model phase functions of \citet{Cahoy2010a, Madhusudhan2012, Greco2015}. This is \replaced{somewhat expected}{not unexpected} since \replaced{the former's}{these} models \added{all} assume \replaced{no clouds}{idealized uniform, scattering atmospheres}. In reality, we know \deleted{is} Jupiter is \replaced{somewhere in between}{more complex}. \added{The \citet{Cahoy2010a} phase functions of a 3x solar metallicity Jupiter-like exoplanet through the Cassini filter set are shown together in \autoref{fig:cahoy} along with the residuals when comparing the model to the data. The models lack a cusp effect and demonstrate a steep slope in filters and there is a small evolution from shorter to longer wavelengths where the bluer wavelengths have a slightly shallower fall off with phase angle before falling off more quickly at phase angles beyond $150^\circ$.} Despite the \citet{Cahoy2010a} assumptions, \added{the residuals show that} their models do a reasonable job at matching Jupiter's phase function \added{especially in the reddest Cassini/ISS filters where the presence of haze is minimal and clouds are relatively uniform.}\deleted{. In wavelengths where the presence of haze is minimal and clouds are relatively uniform, replaced{namely not the VIO filter}{namely the reddest Cassini/ISS filters}, the Cahoy2010 models provide a decent match to the data. The Cahoy2010 phase functions of a 3x solar metallicity Jupiter-like exoplanet through the Cassini filter set are shown together in fig:cahoy. The lack of a cusp effect and steep slope are present in all filters and there is a small evolution from shorter to longer wavelengths where the bluer wavelengths have a slightly shallower fall off with phase angle before falling off more quickly at phase angles beyond $150^\circ$.}

\begin{figure}
\centering
\includegraphics[width=\columnwidth]{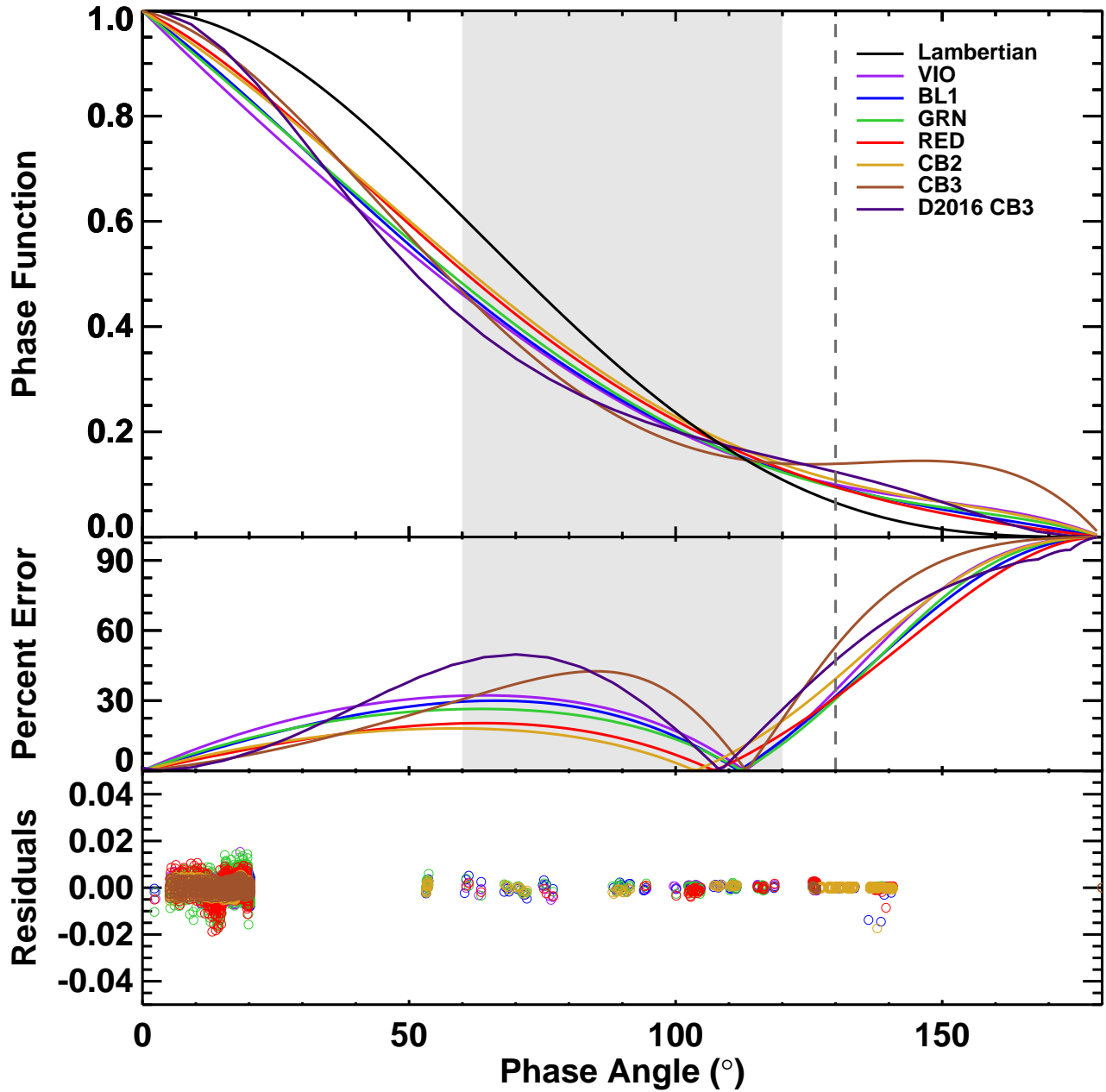}
	\caption{\label{fig:polys} Upper: The {\polyorder} order polynomial fits to each phase function. The shaded region indicates the phase-angles most likely to be observed by direct-imaging missions, 60--120 degrees. Note that the fits are unconstrained by the data past \replaced{$\sim145^\circ$ phase}{$\sim130^\circ$ phase as marked by the vertical dashed line}. \replaced{Lower:}{Middle:} The percent \replaced{error}{difference} between the Lambertian phase function and each filter. \added{Lower: The residuals between the data and each fit.}}
\end{figure}

The polynomial fits to the Cassini data are shown all together in the upper panel of \autoref{fig:polys} in comparison with a Lambertian phase function. In the Cassini data, CB3 shows the weakest demonstration of the cusp effect \replaced{and some argument can be made for its absence in BL1 and CB2. However, even those filters do not align particularly well with any of the models}{and yet, as demonstrated by \citet{Dyudina2016}, still does not align well with any of the models}. The mid phase angle slope of the data is shallower than any of the models. The gray shaded region from 60--120 degrees highlights the phase angles where \added{WFIRST} observations will most often occur. The gray dashed vertical line indicates where our fits become unconstrained by the data and any behavior beyond this point should be treated with caution.

For future direct-imaging missions, the desire for a bright planet that is still widely separated provides constraints on what the optimal phase angles of observations might be. The planet/star contrast ratios for the WFIRST candidate systems are computed by \citet{WFIRST} at $60^\circ$. \autoref{fig:polys} shows that in the near-quadrature regime Jupiter-like planets are much darker than predicted by the Lambertian assumption. The \replaced{lower panel}{middle} panel of \autoref{fig:polys} shows the percent error ($100|Y-X|/X$, where $Y$ is the predicted value and $X$ is the actual value at each phase) between the Lambertian phase function and Jupiter in each filter. The difference is largest at $60^\circ$ phase where the Lambertian overpredicts Jupiter's brightness by on average roughly 25\%. Meanwhile, \replaced{an observation of a planet near $110^\circ$ phase may be indistinguishable from a Lambertian object}{observations of Jupiter only near $110^\circ$ phase would result in the erroneous conclusion that Jupiter is consistent with a Lambertian sphere}. The Lambertian assumption is a poor model of Jupiter-like exoplanets and further study of the Solar System giants is needed.

\begin{figure}
\centering
\includegraphics[width=\columnwidth]{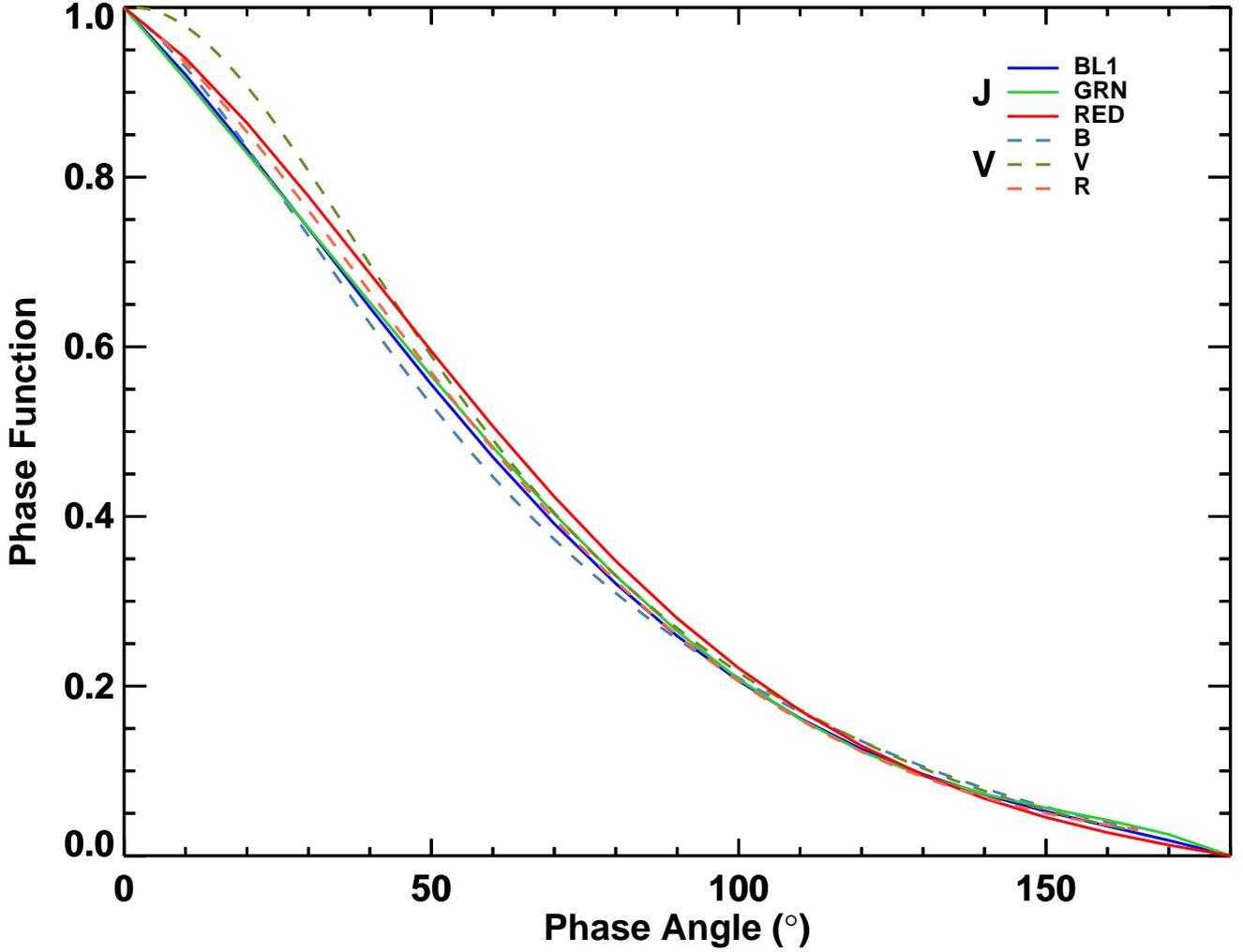}
	\caption{\label{fig:venus} Comparison of phase functions for Venus and Jupiter. The Jupiter curves (solid) are from this work, while those from Venus (dashed) are obtained from \citet{Mallama2006}. The three filters for Jupiter are the closest matching available to the JC system's BVR.}
\end{figure}

\added{Given the complexity of Jupiter's true phase curve, we briefly consider another Solar System planet, Venus, which has a well studied atmosphere. Venus is interior to the orbit of the Earth and its orbit has allowed for extensive ground- and space-based monitoring of its brightness through time. \citet{Mallama2006} provide polynomial fit coefficients to Venus' phase curves which are valid from $\sim$6 to $\sim$165 degrees in the JC system. We computed the $0^\circ$ phase angle point for normalization and demonstrate the differences between the two planets in \autoref{fig:venus}. Although WFIRST will not be able to observe Venus-like planets, the next generation of space-based telescopes may well be able to. We note that like Jupiter, Venus' phase curve differs from that of a simple Lambertian and the interpretation of observations of such worlds will likewise require sophisticated modeling such as that of \citet{Robinson2010}.}

As discussed previously, there is some discussion over optimal color index choices. The desire to measure the blue continuum slope and the strength of the redder lines indicates a color index selection of $(B - V)$, $(V - Z)$ or $(V - I)$. \citet{Cahoy2010a} preferred similar indices, $(B - V)$, $(R - I)$ and $(B - V)$, $(V - I)$, and \citet{Hegde2013} preferred $(B - I)$, $(B - V)$. For comparison purposes, we have \replaced{used}{explored this question using} the Cassini/ISS filters with the closest effective wavelengths, namely BL1 is representative of the B filter, GRN is V, RED is R, CB2 is I and CB3 is Z. 

To compute color, we use the polynomial fit functions scaled to align with \citet{Karkoschka1998} albedo values, $P_K$, and compute them at 1 degree increments from 0--135 degrees. We then multiply by the filter-integrated value of the solar reference spectrum at 5.2~AU that was previously divided out by CISSCAL during the reduction process. The color indices are computed as shown in \autoref{eqn:index} for two filters, $f_1$ and $f_2$.

\begin{equation}
\label{eqn:index}
f_1-f_2 = -2.5\log \left( \frac{P_{K,f_1}(\alpha)\int_\lambda F_\odot(\lambda)T_{f_1}(\lambda)d\lambda}{P_{K,f_2}(\alpha)\int_\lambda F_\odot(\lambda)T_{f_2}(\lambda)d\lambda}\right)
\end{equation}

\begin{figure*}
\centering
\includegraphics[width=\linewidth]{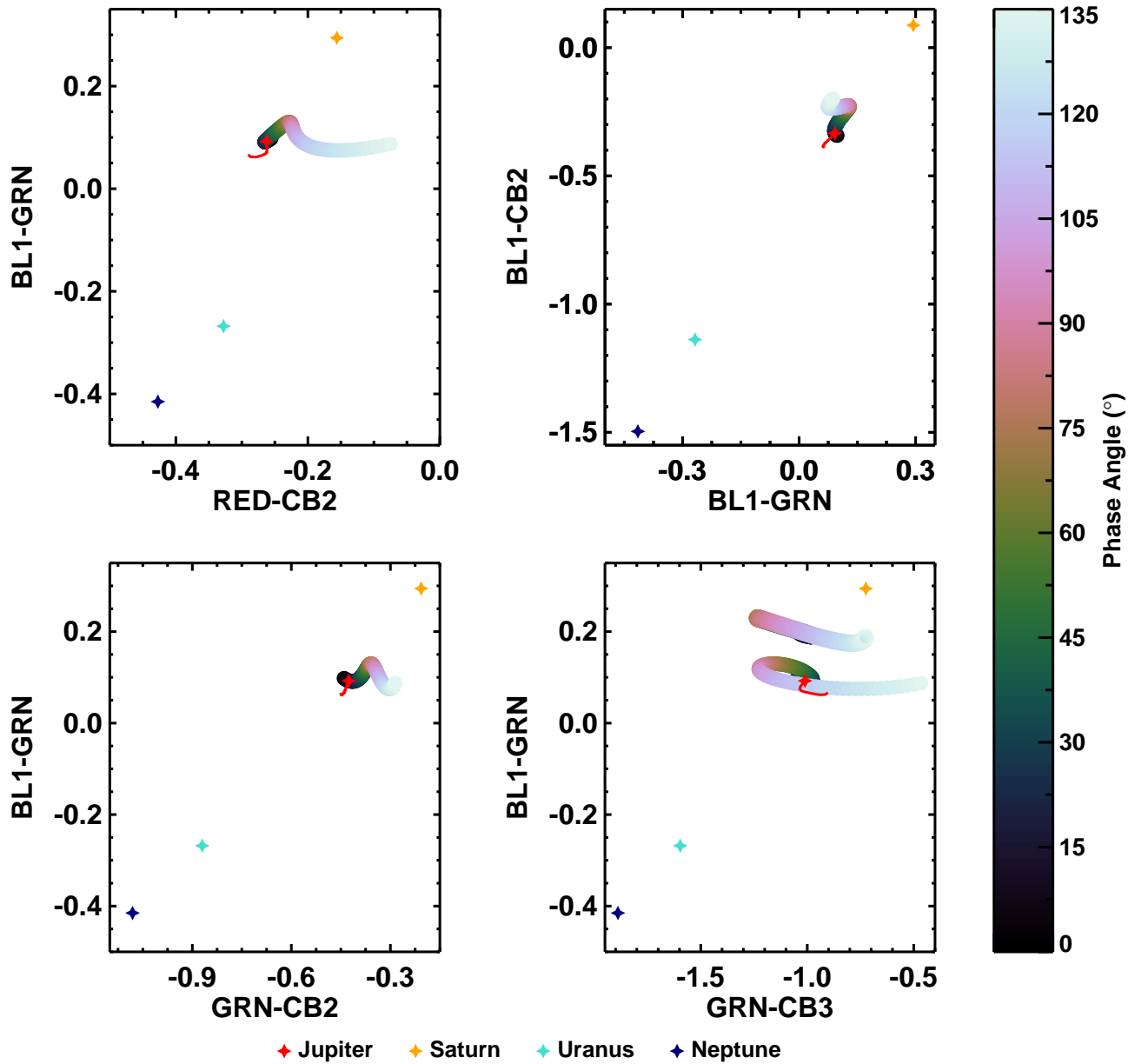}
\caption{\label{fig:color}Color-color diagrams for Jupiter as a function of phase angle for a variety of color indices. For reference, the colored stars indicate the \citet{Karkoschka1998} filter-integrated color at $6.8^\circ$ phase for Jupiter in red, $5.7^\circ$ phase for Saturn in orange, $\sim0^\circ$ phase for Uranus in turquoise, and $\sim0^\circ$ phase for Neptune in navy. The 3x solar metallicity Jupiter-like planet of \citet{Cahoy2010a} is shown completely in red tracks and is much less variable than Jupiter. \added{For the panel including CB3, we also demonstrate the variation assuming the \citet{Dyudina2016} model offset by +0.1 in BL1-GRN for clarity.}}
\end{figure*}

The color-color diagrams as a function of phase angle are shown in \autoref{fig:color} with the colors of all the giant planets as a reference. The colors of all the giant planets were computed from \citet{Karkoschka1998} and represent the colors of those planets at $6.8^\circ$ \added{for Jupiter, $5.7^\circ$ phase for Saturn, and $\sim0^\circ$ phase for Uranus and Neptune}. For comparison, we have also computed the color of the 3x solar metallicity Jupiter-like planet at a separation of 5~AU using the phase function from \citet{Cahoy2010a} also scaled to \citet{Karkoschka1998} and those diagrams are plotted without phase information. 

The $0^\circ$ phase angle colors are not in perfect agreement with \citet{Cahoy2010a}, but are in the correct region of color space. \citet{Cahoy2010a} expressed concern with selecting color indices where the planet types are well separated. \autoref{fig:color} demonstrates that the giant planets can occupy distinct regions of color space. \deleted{The diagrams show that} Jupiter's color varies \replaced{rather wildly throughout an orbit and is not well behaved in these color spaces. This}{with phase angle}. This variation is much larger than predicted by the idealized model of \citet{Cahoy2010a} ranging from $\sim0.1$~mag in (BL1 - GRN) up to \replaced{$\sim0.8$~mag in (GRN - CB3)}{$\sim0.5$~mag in (GRN - CB3) assuming the model by \citet{Dyudina2016}}. \added{A Jupiter-twin may fall somewhere within this much larger range depending on the phase in which it is observed.} This may exacerbate the issue of using exoplanet colors as a method of distinguishing between planet types and background objects. 

The precise evolution of color with phase angle will vary based on filter selection. The \citet{Cahoy2010a} models show less variation in the Cassini/ISS photometric system than in the JC system and the differences in color between the planets is also smaller. \replaced{Given this behavior, it is worth noting despite lack of UVBRI observations of Jupiter that it is likely that the variations will still be proportionately large in that photometric system.}{It is possible that the Cassini observations in the JC system would result in Jupiter's color variations being proportionately larger.}

The Cassini/ISS filters were specifically selected to target known features while the JC system was an earlier evolution of a filter set intended for stellar classification. For future direct-imaging missions, the filter set must also evolve to better allow classification of planets. In particular, observations will most often be done at phase angles between 60 and 120 degrees (olive and lighter in \autoref{fig:color}). The variation in this regime ranges from $\sim0.01$~mag to $\sim0.3$~mag. The WFIRST/CGI filters will not be in the JC system, but the filters listed in \citet{WFIRST} have central wavelengths similar to those of Cassini/ISS's RED, CB2, and MT3 filters \deleted{while the polarized filters are bluer and coincide with BL1, GRN, IR2, and MT3}. These filter choices are intended for measuring the blue continuum slope and the effects of methane and given their similarities to the Cassini/ISS filterset should be well suited to differentiating planets \added{as long as appropriate care is given to measure the phase variation over an adequate fraction of the planetary orbit}.

\section{Conclusions}
\label{sec:conc}
We have presented a study of Jupiter's reflected light as seen by Cassini/ISS during the millennium flyby in late 2000 to early 2001. We have analyzed the full-disk images of Jupiter from both cameras in six filters spanning the range of wavelengths 400--1000~nm and phases from \replaced{0--150}{0--140} degrees.

We find that Jupiter's phase function \added{as observed in 2000 and 2001} is \replaced{best explained}{well reproduced} by a {\polyorder}-order polynomial whose coefficients are given in \autoref{tbl:polycoeffs} for each Cassini/ISS filter. In general, the brightness of Jupiter falls off more steeply near full illumination than previously predicted and is significantly darker than Lambertian near quadrature and somewhat brighter than both a purely Rayleigh scattering model and the \citet{Cahoy2010a} prediction at high phase angles.

We find that Jupiter's color varies \replaced{drastically}{substantially} with phase angle by as much as $\sim0.8$~mag across all phase angles and as much as $\sim0.3$~mag in the 60--120 degree regime, well beyond that previously predicted by \citep{Cahoy2010a}. \deleted{Without proper filter selection, Jupiter-like planets could potentially masquerade as other classes of planets and/or be difficult to differentiate.} Although not demonstrated here, we expect, given the complexity of planetary atmospheres, that such color changes are typical of many types of planets. 

The variability that we observe in Jupiter's color as a function of phase, demonstrates that multi-bandpass measurements would be required to fully characterize a given planet. The phase dependence likely arises from the relative contributions to the scattered light from clouds and hazes at various altitudes in the atmosphere and carries information about the vertical structure and composition of the atmosphere that could not easily be extracted from a single observation.

\deleted{Future direct-imaging missions should carefully target known differentiating features such as the blue continuum slope and the strength of the red features. The WFIRST filters target features similar to those identified by the Cassini/ISS team and this should alleviate some of the classification issues.}

We analyzed only six of the filters on Cassini/ISS and lack of phase coverage prevented the rest from being included in this study. In particular, the methane filters, which lacked the requisite phase coverage to be included in our study, but has been included in others \citep{Dyudina2016}, and the polarization filters are of particular interest to the studies of planet color especially with respect to WFIRST.

\added{As the field of exoplanet science moves toward the direct-imaging of planets in reflected light, observations of solar system planets under similar viewing geometries and at similar wavelengths, as presented here, will undoubtedly provide important context. While the new worlds will inevitably present new surprises, viewing them in the context of solar system observations will provide a framework for interpreting the unexpected.}

\acknowledgements
The authors would like to thank Adam Burrows for providing the code to generate the Rayleigh phase curves. K.R. and M.S.M thank the WFIRST Project Office for support of their participation in this project. L.C.M would like to acknowledge support from the NSF GRFP. This material is based upon work supported by the National Science Foundation Graduate Research Fellowship Program under Grant No. DGE-1144458. Any opinions, findings, and conclusions or recommendations expressed in this material are those of the authors(s) and do not necessarily reflect the views of the National Science Foundation.

\pagebreak
\bibliographystyle{aasjournal}
\bibliography{C:/Users/Laura/Documents/Bibs/Thesis-Jupiter}

\end{document}